\documentstyle[12pt]{article}

\begin{document}

{\it University of Shizuoka}

\hspace*{9.5cm} {\bf US-96-04}\\[-.3in]

\hspace*{9.5cm} {\bf March 1996}\\[-.3in]

\hspace*{9.5cm} {\bf hep-ph/9603376}\\[.3in]


\begin{center}

{\Large\bf  Neutrino Mixing in a }\\[.2in]

{\Large\bf   Democratic-Seesaw-Mass-Matrix Model }\\[.5in]

{\bf Yoshio Koide}\footnote{
E-mail: koide@u-shizuoka-ken.ac.jp} \\

Department of Physics, University of Shizuoka \\ 
52-1 Yada, Shizuoka 422, Japan \\[.5in]

{\large\bf Abstract}\\[.1in]

\end{center}

\begin{quotation}
On the basis of a seesaw-type mass matrix model for quarks and leptons, 
$M_f \simeq m_L M_F^{-1} m_R$, where $m_L\propto m_R$ are universal 
for $f=u,d,\nu$ and $e$ (up-quark-, down-quark-, neutrino- and charged 
lepton-sectors), and $M_F$ has a form [(unit matrix)+(democratic-type 
matrix)],  neutrino masses and mixings are investigated.  
It is tried to understand a large $\nu_\mu$-$\nu_\tau$ mixing, 
i.e., $\sin^2 2\theta_{23}\sim 1$, with 
$m_{\nu 1} \ll m_{\nu 2} \sim m_{\nu 3}$, which has been suggested 
by the atmospheric neutrino data.
\end{quotation}

\newpage

\noindent
{\large\bf 1.\ Introduction}

\vglue.1in

The Kamiokande collaboration [1] has recently  suggested 
a possibility of a large neutrino mixing $\nu_\mu$-$\nu_x$, 
$\sin^2 2\theta \simeq 1$, with 
$\Delta m^2 \simeq 1.8\ (1.6)\, \times 10^{-2}$ eV$^2$ for $x=e$  
($x=\tau$) from their atmospheric neutrino data. 
Although their conclusion is still controversial [2], it seems to be
worth while to take it seriously. 
On the other hand, the solar neutrino data [3] 
with the Mikheyev-Smirnov-Wolfenstein (MSW) effect [4] have suggested
a neutrino mixing $\sin^2 2\theta \simeq 7\times 10^{-3}$ with 
$\Delta m^2\simeq 6\times 10^{-6}$ eV$^2$. 
What is of great interest to us is whether we can give a satisfactory 
explanation of both the data, [1] and [3], on the basis of an 
extension of a successful quark mass matrix model to the neutrino sector.

Recently, based on a seesaw-type quark mass matrix model [5],
Fusaoka and the author [6] have proposed a quark mass matrix model
which can naturally understand the observed facts $m_t \gg m_b$ and 
$m_u \sim m_d$, without bringing such a parameter as 
a parameter in $M_u$ takes extremely large value
compared with that in $M_d$.
They have assumed vector-like heavy fermions $F_i$ in addition to 
conventional quarks and leptons $f_i$ ($i=1,2,3$) [$f=u$ (up-quarks), 
$f=d$ (down-quarks), $f=\nu$ (neutrinos) and $f=e$ (charged leptons)].
These fermions belong to  $F_L=(1,1)$, $F_R=(1,1)$, 
$f_L=(2,1)$, and $f_R=(1,2)$ of SU(2)$_L \times$SU(2)$_R$, 
respectively.
The mass matrix for $(f, F)$ is given by a $6\times 6$ matrix
$$
M=\left(
\begin{array}{ccc}
0 & m_L \\
m_R & M_F
\end{array} \right) =
m_0 \left(
\begin{array}{cc}
0 &  Z \\
\kappa Z & \lambda O_f \\
\end{array}
\right) \ ,   \eqno(1.1)
$$
where  the chiral symmetry breaking terms $m_L$ and $m_R$ are assumed 
to be $m_L\propto m_R$ and they have a universal structure $Z$
for quarks and leptons $f$ $(=u,d,\nu,e)$,
$$
Z=\left(
\begin{array}{ccc}
z_1 & 0 & 0 \\
0 & z_2 & 0 \\
0 & 0 & z_3 
\end{array} \right)  \ , \eqno(1.2)
$$
where $z_i$ are normalized as $z_1^2 +z_2^2 +z_3^2=1$. 
The heavy fermion mass matrix $M_F=m_0 \lambda O_f$ has a structure [7] 
of [(unit matrix)+(a democratic-type matrix)] and   
it includes only one complex parameter $b_f e^{i\beta_f}$
which depends on $f=u,d,\nu, e$:
$$
O_f= {\bf 1} + 3 b_f e^{i\beta_f} X \ , \eqno(1.3)
$$
where ${\bf 1}$ are a $3\times 3$ unit matrix and $X$ is a 
democratic-type matrix [8]
$$
X=\frac{1}{3} \left(
\begin{array}{ccc}

1 & 1 & 1 \\
1 & 1 & 1 \\
1 & 1 & 1 
\end{array} \right) \ . \eqno(1.4)
$$
The  mass matrix (1.1) leads to the well-known seesaw form 
$M_f\simeq m_L M_F^{-1} m_R$
for ${\rm Tr}\, M_F \gg {\rm Tr}\, m_R,  {\rm Tr}\, m_L$.
Note that the inverse matrix of $O_f$  again takes the form 
[(unit matrix)+(democratic-type matrix)], 
$$
O_f^{-1}={\bf 1} + 3 a_f e^{i\alpha_f} X \ , \eqno(1.5)
$$
with
$$
a_f e^{i\alpha_f} =- \frac{b_f e^{i\beta_f}}{1+3b_f e^{i\beta_f}} \ .
\eqno(1.6) 
$$
The limit $b_f e^{i\beta_f} \rightarrow -1/3$ leads to 
$|a_f|\rightarrow \infty$.
Therefore, a slight difference between $b_u$ and $b_d$ around 
$b_f\simeq -1/3$ can induce an extremely large difference between 
$m_t$ and $m_b$.
On the other hand, we can keep $m_u\sim m_d$ because 
the democratic mass matrix makes only the third family heavy. 
Thus, they [6] have given a natural explanation of the observed facts 
$m_t \gg m_b$ and $m_u \sim m_d$. 

In order to fix the parameters $z_i$, 
they have assumed that $b_e=0$, i.e., 
$$
M_e\simeq m_0 \frac{\kappa}{\lambda} Z^2 \ , \eqno(1.7)
$$ 
so that $z_i$ are given by 
$$
\frac{z_1}{\sqrt{m_e}}=\frac{z_2}{\sqrt{m_\mu}}=
\frac{z_3}{\sqrt{m_\tau}}=\frac{1}{\sqrt{m_e+m_\mu+m_\tau}} \ . 
\eqno(1.8) 
$$
By taking $\kappa/\lambda=0.02$, $b_u=-1/3$ ($\beta_u=0$) and 
$b_d=-1$ ($\beta_d=-18^\circ$), they have obtained reasonable quark 
mass ratios and Kobayashi-Maskawa (KM) [9] matrix parameters.

In their model, the variety of the quark and lepton mass matrices 
come form the variety of the corresponding heavy fermion mass matrices
which are characterized by the parameter $b_f e^{i\beta_f}$. 
They have concluded that the parameter values $b_u=-1/3$, 
$b_d=-1$ and $b_e=0$ are favorable 
to the observed mass spectra and mixings. 
However, why the nature chooses such values of $b_f$ is an open question. 
In order to obtain a clue to such a question, 
in the present paper, we investigate what value of $b_\nu$ is required 
from the phenomenological study of neutrino masses and mixings.

\vspace{.2in}

\noindent
{\large\bf 2.\  Neutrino mass matrix }
\vglue.1in

In the model in Ref.~[6], the mass matrices of the charged leptons 
and quarks have been given by (1.1). 
In order to understand why neutrino masses are so negligibly small, 
we must consider that a value of the parameter $\lambda$ in (1.1) 
in neutrino sector takes extremely large value 
compared with those in charged lepton 
and quark sectors, or that a extremely large Majorana mass term causes 
the so-called seesaw mechanism [10] doubly. 
The former case is not natural from the standpoint of  the unified 
description of quark and lepton mass matrices.
For the latter case, two possibilities are considered: 
one is that the heavy neutrinos $N_{Li}$ and $N_{Ri}$ have  
large Majorana masses $M_M$, 
and another is that the right-handed 
neutrinos $\nu_{Ri}$ have  large Majorana masses $M_M$.
Roughly speaking, for Tr$M_M\gg {\rm Tr}M_D$ (for convenience, 
we denote the Dirac masses $M_F$ in (1.1) as $M_D$), 
the former  and latter cases lead to  mass matrices for 
the left-handed neutrinos $\nu_{Li}$, 
$$
M_{\nu_L}\simeq -(1/2)^2 m_L M_M^{-1} m_L^T \ , \eqno(2.1)
$$ 
and 
$$
M_{\nu_L}\simeq -(1/2)^4 m_L M_D^{-1} m_R M_M^{-1} 
m_R^T (M_D^T)^{-1} m_L^T 
\ , \eqno(2.2)
$$
respectively.
In the former case, in order to give neutrino mixings, we must consider 
some structure of $M_M$, which may be independent of that of $M_D$,  
so that the mass matrix $M_{\nu_L}$ cannot be related to the 
mass matrices of charged leptons and quarks. 
In the present paper, we investigate the latter possibility.

The $6\times 6$ mass matrix which is sandwiched by 
$(\overline{\nu}_L, \overline{\nu}_R^c, \overline{N}_L, \overline{N}_R^c)$ 
and $(\nu_L^c, \nu_R, N_L^c, N_R)^T$ is given by
$$
M = \left(
\begin{array}{cccc}
0 & 0 & 0 & \frac{1}{2}m_L \\
0 & M_M & \frac{1}{2}m_R^T & 0 \\
0 & \frac{1}{2}m_R & 0 & M_D \\
\frac{1}{2}m_L^T & 0 & M_D^T & 0 
\end{array} \right) \ , \eqno(2.3)
$$
so that the $3\times 3$ light-neutrino mass matrix 
is given by (2.2). 
We assume that $M_M$ is simply given by $M_M=m_0 \xi {\bf 1}$, while 
$M_D$ is given by a universal structure $M_D=m_0 \lambda O_f
=m_0 \lambda ({\bf 1}+3 b_\nu e^{i\beta_\nu}X)$ 
as well as those in quark sectors.
Then, we obtain 
$$
M_{\nu_L} \simeq \frac{1}{16} \frac{\kappa^2 m_0}{\lambda^2 \xi}
Z O_\nu^{-1} Z \cdot Z O_\nu^{-1}Z  \ . \eqno(2.4)
$$

In Fig.~1, we illustrate  the behavior of the neutrino masses versus 
the parameter $b_\nu$, 
which is similar to that of the quark masses (see Fig.~1 in Ref.~[6]).
For the case of $\beta_\nu=0$, 
at $b_\nu=-1/2$ ($b_\nu=-1$), the mass levels $m_{\nu_3}$ and 
$m_{\nu_2}$ ($m_{\nu_2}$  and $m_{\nu_1}$) degenerate each other. 
Therefore, we can  expect that large neutrino mixings 
occur at $b_\nu=-1/2$ and $b_\nu=-1$. 
For the case of $\beta_\nu \neq 0$, the degeneracies between $m_{\nu_i}$ 
and $m_{\nu_j}$ disappear, so that the large mixings 
$\sin^2 2\theta_{ij}\simeq 1$  become mild.

\vspace{.2in}

\noindent
{\large\bf 3. Masses and mixings for typical three cases of $b_\nu$}

Let us show the neutrino masses $m_i$ and mixing matrix $U_{\nu_L}$ for 
typical three cases of $b_\nu$: 
$b_\nu\simeq -1/3$, $b_\nu \simeq -1/2$ and $b_\nu \simeq -1$. 
Here, the mixing matrix $U_{\nu_L}$ is defined by 
$$
\nu_\alpha = \sum_{i=1}^3\left( U_{\nu_L}\right)_{\alpha i} \nu_i 
\ , \eqno(3.1)
$$
where $\nu_\alpha$ ($\alpha=e, \mu, \tau$) are flavor eigenstates and 
$\nu_i$ ($i=1,2,3$) are mass eigenstates. 
For simplicity, we consider the case of $\beta_\nu=0$. 
Then, we obtain the following approximate expressions:
$$ 
m_{\nu_1}\simeq \left( \frac{3}{4} \frac{m_e}{m_\tau}\right)^2 m_0^\nu\ , 
\ \ \ m_{\nu_2}\simeq \left(\frac{ m_\mu}{m_\tau}\right)^2
m_0^\nu \ , \ \ \ m_{\nu_3}\simeq \left(\frac{\sqrt{2}}{27
|\varepsilon|}\right)^2 m_0^\nu
 \ ,  \eqno(3.2)
$$
$$
U_{\nu L} \simeq \left(
\begin{array}{ccc}
1 & \frac{1}{2}\sqrt{m_e/m_\mu} & \sqrt{m_e/m_\tau} \\
-\frac{1}{2}\sqrt{m_e/m_\mu} & 1 & \sqrt{m_\mu/m_\tau} \\
-\frac{1}{2}\sqrt{m_e/m_\tau} & -\sqrt{m_\mu/m_\tau} & 1 
\end{array} \right) \ ,\eqno(3.3)
$$
for $b_\nu\simeq -1/3$ ($\varepsilon\equiv b_\nu +1/3$),
$$ 
m_{\nu_1}\simeq \left( \frac{m_e}{m_\tau} \right)^2m_0^\nu 
\ , \ \ \ 
m_{\nu_2}\simeq m_{\nu_3} \simeq \left(
\frac{1}{2} \sqrt{\frac{m_\mu}{m_\tau}} \right)^2m_0^\nu 
\ ,  \eqno(3.4)
$$
$$
U_{\nu L} \simeq \left(
\begin{array}{ccc}
1 & \frac{1}{\sqrt{2}}\left(\sqrt{\frac{m_e}{m_\mu}}+\eta 
\sqrt{\frac{m_e}{m_\tau}}\right) & 
\frac{1}{\sqrt{2}}\left(\sqrt{\frac{m_e}{m_\mu}} -\eta 
\sqrt{\frac{m_e}{m_\tau}}\right)  \\
 -\sqrt{m_e/m_\mu} & \frac{1}{\sqrt{2}} & 
-\eta\frac{1}{\sqrt{2}} \\
-\sqrt{m_e/m_\tau} & \eta\frac{1}{\sqrt{2}} & \frac{1}{\sqrt{2}} \\
\end{array} \right) \ ,\eqno(3.5)
$$
for  $b_\nu \simeq -1/2$, and 
$$ 
m_{\nu_1}\simeq m_{\nu_2}\simeq  \left(
\frac{1}{2} \sqrt{\frac{m_e m_\mu}{m_\tau^2}}\right)^2m_0^\nu \ , \ \ \ 
m_{\nu_3}= \left(\frac{1}{4} \right)^2m_0^\nu \ ,  \eqno(3.6)
$$
$$
U_{\nu L} \simeq \left(
\begin{array}{ccc}
\frac{1}{\sqrt{2}} & -\eta\frac{1}{\sqrt{2}} & -\sqrt{m_e/m_\tau}  \\
\frac{1}{\sqrt{2}} & \frac{1}{\sqrt{2}} & 
-\sqrt{m_\mu/m_\tau} \\
\eta\frac{1}{\sqrt{2}} \left( \sqrt{\frac{m_\mu}{m_\tau}}
+\eta\sqrt{\frac{m_e}{m_\tau}}\right) &
\frac{1}{\sqrt{2}} \left( \sqrt{\frac{m_\mu}{m_\tau}}
-\eta\sqrt{\frac{m_e}{m_\tau}}\right) &  1 
\end{array} \right) \ ,\eqno(3.7)
$$
for $b_\nu \simeq= -1$, where $m_0^\nu$ is defined by 
$$
m_0^\nu = \left(\frac{\kappa}{2\lambda}\right)^2 \frac{m_0}{\xi} 
\ . \eqno(3.8)
$$
Here, in (3.5) [(3.7)], the factor $\eta$ is defined as $\eta=\pm 1$ 
for $b_\nu=b^0_{23}\mp \varepsilon \simeq -1/2$ ($1\gg\varepsilon>0$) 
[$b_\nu=b^0_{12}\pm \varepsilon \simeq -1$ ($\varepsilon>0$)], 
where $b^0_{23}$ [$b^0_{12}$] is the value of $b_\nu$ at which the masses 
of $\nu_2$ and $\nu_3$ [$\nu_1$ and $\nu_2$] exactly degenerate. 
As shown in (3.5) and (3.7), the mixing elements $(U_{\nu L})_{\alpha 2}$ 
and $(U_{\nu L})_{\alpha 3}$  [$(U_{\nu L})_{\alpha 1}$ and 
$(U_{\nu L})_{\alpha 2}$ ] are exchanged each other at $b_\nu=b^0_{23}$ 
[$b_\nu=b^0_{12}$], because the mass levels of $\nu_2$ and $\nu_3$ 
[$\nu_1$ and $\nu_2$] cross each other at   $b_\nu=b^0_{23}$ 
[$b_\nu=b^0_{12}$] as  seen in Fig.~1. 

The result (3.3) for the case $b_\nu\simeq -1/3$ has been reported in 
Ref.~[11]. 
The mixing matrix element $U_{e 2}\equiv \sin\theta_{e 2}$ leads to 
$\sin^2 2\theta_{e 2}\simeq m_e/m_\mu = 4.8 \times 10^{-3}$, 
which is in good agreement with the MSW solution of solar neutrino data 
[3] $\sin^2 2\theta \simeq 7\times 10^{-3}$.
However, in this paper, we will direct our attention to the atmospheric 
neutrino data [1] as well as the solar neutrino data [3].

\vspace{.2in}

\noindent
{\large\bf 4. Numerical study}
\vglue.1in

We consider that the atmospheric neutrino data [1] show 
$\nu_\mu$-$\nu_\tau$ mixing, while the solar neutrino data [3] show 
$\nu_e$-$\nu_\mu$ mixing.

For reference, in Fig.~2, 
we illustrate $\Delta m^2_{21}\equiv m^2_{\nu_2}-m^2_{\nu_1}$ 
versus $\sin^2 2\theta_{e2}\equiv 4|U_{e2}|^2(1-|U_{e2}|^2)$ and  
$\Delta m^2_{32}\equiv m^2_{\nu_3}-m^2_{\nu_2}$ 
versus $\sin^2 2\theta_{\mu 3}\equiv 4|U_{\mu 3}|^2
(1-|U_{\mu 3}|^2)$ in the case of $\beta_\nu=0$.
Note that the value of $\sin^2 2\theta_{e 2}$ is discontinuous at 
$b_\nu \simeq -1/2$ because the value of $b_\nu$ crosses the value 
$b_{23}^0\simeq -1/2$.

We interests in the case of 
$b_\nu \simeq -1/2$, because the case yields 
$\sin^2 2 \theta_{\mu 3} \sim 1$ 
with $m_{\nu_1} \ll m_{\nu_2} \simeq m_{\nu_3}$.
In Fig.~3, we illustrate the behaviors of $\sin^2 2\theta_{e2}$ and 
$\sin^2 2\theta_{\mu 3}$ versus $b_\nu$. 
For reference, we also illustrate the ratio 
$\Delta m^2_{32}/\Delta m^2_{21}$ in the figure.
The observed values $\Delta m^2_{32}\simeq 1.6 \times 10^{-2}$ eV$^2$ [1] 
and $\Delta m^2_{21}\simeq 6 \times 10^{-6}$ eV$^2$ [3]
give the ratio $\Delta m^2_{32}/\Delta m^2_{21}\simeq 2.7 \times 10^3$.
As seen in Fig.~3, there is no solution which gives 
$\sin^2 2\theta_{\mu 3}\simeq 1$,  $\sin^2 2\theta_{e 2}\simeq 0.007$ 
and $\Delta m^2_{32}/\Delta m^2_{21}\simeq 3\times 10^3$ 
simultaneously. 
If we reduce the requirement of the maximal mixing 
$\sin^2 2\theta_{\mu 3}\simeq 1$, for example, to 
$\sin^2 2\theta_{\mu 3}\simeq 0.4$, we can find satisfactory 
solutions of $b_\nu$.
For the case of $\beta_\nu=0$, the choice $b_\nu \simeq -0.41$ can 
give the plausible values of $\sin^2 2\theta_{\mu 3}$, 
$\sin^2 2\theta_{e 2}$ and $\Delta m^2_{32}/\Delta m^2_{21}$ 
as seen in Fig.~3. 
For the case of $\beta_\nu \neq 0$, we take $b_\nu=-1/2$ 
by way of trial, because the value is a simple fractional number
which gives $b_\nu \sim -0.5$. 
Then, the choice $\beta_\nu\simeq 22^\circ$ can give favorable predictions. 
We list numerical results for some special cases of
$(b_\nu, \beta_\nu)$ in Table 1.  

In Table 1, the values $\xi m_0$ have been estimated as follows: from 
(1.7), we obtain 
$$
m_0 \kappa /\lambda = m_\tau +m_\mu +m_e =1.883\ {\rm GeV} \ , \eqno(4.1)
$$
so that from the definition (3.8), we obtain
$$
\xi m_0 =(m_\tau +m_\mu +m_e )^2/4 m_0^\nu \ . \eqno(4.2)
$$
Here, the values of $m_0^\nu$ have been obtained from 
$(\Delta m_{21}^2)_{theory}/(\Delta m_{21}^2)_{input}$ with 
$(\Delta m_{21}^2)_{input}=6\times 10^{-6}$ eV$^2$. 
We find that the Majorana masses of $\nu_R$ are of the order of $10^9$ GeV.

\vspace{.2in}

\noindent
{\large\bf 5. Discussions}
\vglue.1in

As seen in Fig.~3 and Table 1, if we want a solution which gives 
the largest possible $\nu_\mu$-$\nu_\tau$ mixing with 
$\Delta m_{32}^2 \geq 10^{-2}$ eV$^2$ 
(for the input $\Delta m_{21}^2=6\times 10^{-6}$ eV$^2$), 
the solution $b_\nu=-0.41$ with $\beta_\nu=0$ is favorable rather than
the case of $\beta\neq 0$: the mixing matrix $U_{\nu_L}$ and neutrino masses 
$m_{\nu_i}$ are given by 
$$
U_{\nu_L} = \left( 
\begin{array}{rrr}
0.9988 & 0.0387 & 0.0310 \\
-0.0484 & 0.9061 & 0.4203 \\
-0.0117 & -0.4212 & 0.9069 
\end{array} \right) \ , \eqno(5.1)
$$
$m_{\nu_1}=2.4\times 10^{-8}$ eV,  $m_{\nu_2}=0.0024$ eV and  
$m_{\nu_3}=0.099$ eV,  respectively

However, from the phenomenological study [6] of quark masses and KM mixings, 
we have known that the values $b_u=-1/3$ and $b_d=-1$ for the input 
$b_e=0$ are favorable.
If we take notice of an empirical rule that $(b_e, Q_e)=(0, -1)$, 
$(b_d, Q_d)=(-1, -1/3)$ and $(b_u, Q_u)=(-1/3, +2/3)$, where 
$Q_f$ is the charge of the fermions $f_i$, we can speculate [12] 
$(b_\nu, Q_\nu)=(+2/3, 0)$ for the neutrino sector.
The value $b_\nu=2/3$ with $\beta_\nu=0$ ($\beta_\nu=\pi$) 
predicts $\sin^2 2\theta_{e2}=3.2\times 10^{-3}$ $(0.074)$, 
$\sin^2 2\theta_{\mu 3}=0.021$ $(0.52)$ and 
$\Delta m^2_{32}/\Delta m^2_{21}=1.2\times 10^5$ ($4.1\times 10^3$). 
The predicted values of $\sin^2 2\theta_{\mu 3}$ and 
$\Delta m^2_{32}/\Delta m^2_{21}$ in the case of $(b_\nu, \beta_\nu)=
(+2/3, \pi)$ [i.e., $(b_\nu, \beta_\nu)=(-2/3,0)$] are favorable 
to the observed data, but the predicted value 
$\sin^2 2\theta_{e2}=0.074$ is larger 
by one order than the the MSW-suggested value 
$\sin^2 2\theta_{e2}\simeq 7\times 10^{-3}$.
If we suppose $b_\nu=2/3$ with $\beta_\nu\simeq \pi$, we must discard 
the neutrino mixing $\sin^2 2\theta_{e 2}\simeq 7\times 10^{-3}$ 
with $\Delta m^2_{21}\simeq 6\times 10^{-6}$ eV$^2$, which is 
suggested from the solar neutrino data. 
On the other hand, if we suppose $b_\nu=2/3$ with $\beta_\nu\simeq 0$, 
we must discard the neutrino mixing 
$\sin^2 2\theta_{\mu 3}\sim 1$ with 
$\Delta m^2_{32}\simeq 1.6\times 10^{-2}$ eV$^2$, which is 
suggested from the atmospheric neutrino data. 
If we want an explanation both for the atmospheric and solar neutrino 
data, we must accept the choice $(b_\nu, \beta_\nu)\simeq (-0.41, 0)$, 
but it is an open question how we understand the parameter value 
$b_\nu\simeq -0.41$ with $\beta_\nu \simeq 0$ from the point of view of 
a unified description of $b_f$ $(f=\nu, e, u, d)$.

\vspace{.3in}


\centerline{\large\bf Acknowledgments}

The authors would like to express their sincere thanks to Professors 
R.~Mohapatra, A.~Yu.~Smirnov, and H.~Minakata 
for  their valuable comments 
on a preliminary version of the present work.
He would also like to thank Professor H.~Fusaoka for helpful 
conversations on the democratic seesaw-mass-matrix model.
This work was supported by the Grant-in-Aid for Scientific Research, the 
Ministry of Education, Science and Culture, Japan (No.06640407).

\vspace{.3in}
\newcounter{0000}
\centerline{\large\bf References }
\begin{list}
{[~\arabic{0000}~]}{\usecounter{0000}
\labelwidth=0.8cm\labelsep=.05cm\setlength{\leftmargin=0.7cm}
{\rightmargin=.2cm}}
\item Y.~Fukuda {\it et al}, Phys.~Lett. {\bf B335} (1994) 237.

Also see, Soudan-2 collaboration, M.Goodman {\it et al.}, 
Nucl.~Phys. (Proc.~Suppl.) {\bf B38} (1995) 337; 

IMB collaboration, D.~Casper {\it  et al}, Phys.~Rev.~Lett. {\bf 66} (1989) 
2561; R.~Becker-Szendy {\bf et al}, Phys.~Rev. {\bf D46} (1989) 3720.
\item NUSEX collaboration, M.~Aglietta {\it at al.}, Europhys.~Lett. {\bf 8} 
(1989) 611;

Frejus collaboration, Ch.~Berger {\it el al.}, Phys.~Lett. {\bf B227} (1989)
489; {\it ibid} {\bf B245} (1990) 305; K.Daum {\it et al.}, Z.~Phys. 
{\bf C66} (1995) 417.

\item GALLEX collaboration, P.~Anselmann {\it et al}, 
Phys.~Lett. {\bf B327}  (1994) 377; {\bf B357} (1995) 237;

SAGE collaboration, J.~N.~Abdurashitov {\it et al}, 
Phys.~Lett. {\bf B328}, 234 (1994). 

Also see, N.~Hata and P.~Langacker, Phys.~Rev. {\bf D50} (1994) 632; 
{\bf D52} (1995) 420.
\item S.~P.~Mikheyev and A.~Yu.~Smirnov, Yad.~Fiz. {\bf 42} 
(1985) 1441; [Sov.~J.~Nucl.~Phys. {\bf 42} (1985) 913]; 
Prog.~Part.~Nucl.~Phys. {\bf 23} (1989) 41; 

L.~Wolfenstein, Phys.~Rev. {\bf D17} (1978) 2369; {\bf D20} (1979) 
2634;

T.~K.~Kuo and J.~Pantaleon, Rev.~Mod.~Phys. {\bf 61} (1989) 937. 

Also see, A.~Yu.~Smirnov, D.~N.~Spergel and J.~N.~Bahcall, 
Phys.~Rev. {\bf D49} (1994) 1389.
\item Z.~G.~Berezhiani, Phys.~Lett. {\bf 129B} (1983)  99;
Phys.~Lett. {\bf 150B} (1985)  177;

D.~Chang and R.~N.~Mohapatra, Phys.~Rev.~Lett. {\bf 58} 
1600 (1987); 

A.~Davidson and K.~C.~Wali, Phys.~Rev.~Lett. {\bf 59} (1987)  393;

S.~Rajpoot, Mod.~Phys.~Lett. {\bf A2} (1987)  307; 
Phys.~Lett. {\bf 191B} (1987)  122; Phys.~Rev. {\bf D36} (1987)  1479;

K.~B.~Babu and R.~N.~Mohapatra, Phys.~Rev.~Lett. {\bf 62}  (1989) 1079; 
Phys.~Rev. {\bf D41} (1990)  1286;

S.~Ranfone, Phys.~Rev. {\bf D42} (1990)  3819; 

A.~Davidson, S.~Ranfone and K.~C.~Wali, Phys.~Rev. 
{\bf D41} (1990)  208; 

I.~Sogami and T.~Shinohara, Prog.~Theor.~Phys. {\bf 66}  (1991)  1031;
Phys.~Rev. {\bf D47}  (1993)  2905; 

Z.~G.~Berezhiani and R.~Rattazzi, Phys.~Lett. {\bf B279}  (1992)  124;

P.~Cho, Phys.~Rev. {\bf D48} (1994)  5331; 

A.~Davidson, L.~Michel, M.~L,~Sage and  K.~C.~Wali, Phys.~Rev. 
{\bf D49} (1994)  1378; 

W.~A.~Ponce, A.~Zepeda and R.~G.~Lozano, Phys.~Rev. {\bf D49} 
(1994) 4954.
\item  Y.~Koide and H.~Fusaoka, US-95-03 and AMU-95-04 (1995) 
(hep-ph/9505201), to be published in Z.~Phys.~C (1996).
\item H.~Terazawa, University of Tokyo, Report No.~INS-Rep.-298 
(1977) (unpublished); Genshikaku Kenkyu (INS, Univ.~of Tokyo) 
{\bf 26} (1982) 33.
\item H.~Harari, H.~Haut and J.~Weyers, Phys.~Lett. {\bf B78}
 (1978) 459;

T.~Goldman, in {\it Gauge Theories, Massive Neutrinos and 
Proton Decays}, edited by A.~Perlumutter (Plenum Press, New York, 
1981), p.111;

T.~Goldman and G.~J.~Stephenson,~Jr., Phys.~Rev. {\bf D24} 
(1981) 236; 

Y.~Koide, Phys.~Rev.~Lett. {\bf 47} (1981) 1241; Phys.~Rev. 
{\bf D28} (1983) 252; {\bf 39} (1989) 1391;

C.~Jarlskog, in {\it Proceedings of the International Symposium on 
Production and Decays of Heavy Hadrons}, Heidelberg, Germany, 1986
edited by K.~R.~Schubert and R. Waldi (DESY, Hamburg), 1986, p.331;

P.~Kaus, S.~Meshkov, Mod.~Phys.~Lett. {\bf A3} (1988) 1251; 
Phys.~Rev. {\bf D42} (1990) 1863;

L.~Lavoura, Phys.~Lett. {\bf B228} (1989) 245; 

M.~Tanimoto, Phys.~Rev. {\bf D41} (1990)  1586;

H.~Fritzsch and J.~Plankl, Phys.~Lett. {\bf B237} (1990) 451; 

Y.~Nambu, in {\it Proceedings of the International Workshop on 
Electroweak Symmetry Breaking}, Hiroshima, Japan, (World 
Scientific, Singapore, 1992), p.1.
\item M.~Kobayashi and T.~Maskawa, Prog.~Theor.~Phys. {\bf 49} 
(1973) 652.
\item M.~Gell-Mann, P.~Rammond and R.~Slansky, in {\it Supergravity}, 
edited by P.~van Nieuwenhuizen and D.~Z.~Freedman (North-Holland, 
1979); 

T.~Yanagida, in {\it Proc. Workshop of the Unified Theory and 
Baryon Number in the Universe}, edited by A.~Sawada and A.~Sugamoto 
(KEK, 1979); 

R.~Mohapatra and G.~Senjanovic, Phys.~Rev.~Lett. 
{\bf 44} (1980)  912.
\item Y.~Koide, Mod.~Phys.~Lett. {\bf A8} (1993) 2071.
\item  Y.~Koide and H.~Fusaoka, US-96-02 and AMU-96-01 (1996) 
(hep-ph/9602303).
\end{list}

\newpage

Table 1. Numerical results for special cases of $(b_\nu, \beta_\nu)$.
The input value $\Delta m_{21}^2\equiv 6\times 10^{-6}$ eV$^2$ 
is used in order to fix the value of $m_0^\nu$.

$$
\begin{array}{|c|c|c|c|c|}\hline
(b_\nu, \beta_\nu)   & (-0.41, 0^\circ) &(-0.40, 0^\circ)  
& (-1/2, 20^\circ) & (-1/2, 22^\circ) \\ \hline
\Delta m^2_{21} & 6 \times 10^{-6}\ {\rm eV^2}
& 6 \times 10^{-6}\ {\rm eV^2} & 6 \times 10^{-6}\ {\rm eV^2} 
& 6 \times 10^{-6}\ {\rm eV^2} \\ 
\sin^2 2 \theta_{e2} & 6.1 \times 10^{-3}  & 5.9 \times 10^{-3} 
 & 1.4 \times 10^{-2} & 1.4 \times 10^{-2} \\ 
\hline
\Delta m^2_{32} & 0.97 \times 10^{-2}\ {\rm eV^2} &
2.7 \times 10^{-2}\ {\rm eV^2} & 0.65 \times 10^{-2}\ {\rm eV^2} 
& 1.1 \times 10^{-2}\ {\rm eV^2} \\ 
\sin^2 2 \theta_{\mu 3} & 0.58 & 0.52 & 0.49 & 0.41 \\ \hline
m(\nu_1) & 2.4 \times 10^{-8}\ {\rm eV}
 & 2.6 \times 10^{-8}\ {\rm eV}  & 7.4 \times 10^{-8}\ {\rm eV} 
 & 8.2 \times 10^{-8}\ {\rm eV} \\
m(\nu_2) & 2.4 \times 10^{-3}\ {\rm eV}
 & 2.4 \times 10^{-3}\ {\rm eV}  & 2.4 \times 10^{-3}\ {\rm eV} 
 & 2.4 \times 10^{-3}\  {\rm eV} \\
m(\nu_3) & 0.099\ {\rm eV} & 0.16\ {\rm eV} & 0.081\ {\rm eV} 
& 0.103\ {\rm eV} \\ \hline
m_0^\nu & 0.46 \ {\rm eV}  & 0.50 \ {\rm eV} 
 & 1.25 \ {\rm eV} & 1.44 \ {\rm eV} \\ \hline
\xi m_0 & 1.9\times 10^9 \ {\rm GeV} & 1.8\times 10^9 \ {\rm GeV} & 
0.71\times 10^9 \ {\rm GeV} &  0.62\times 10^9 \ {\rm GeV} \\ \hline 
\end{array}
$$

\vglue.4in

\begin{center}
{\large\bf Figure Captions}
\end{center}

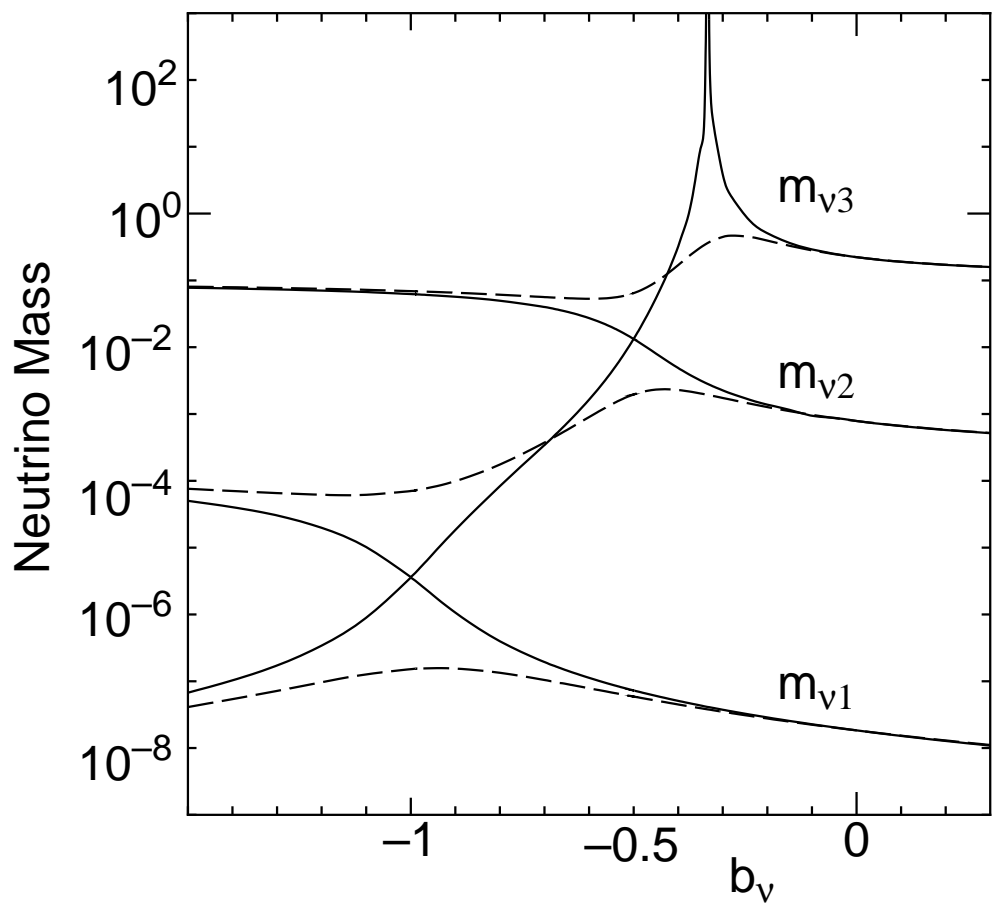
Fig.~1.  Neutrino masses (in unit of $m_0^\nu$) 
versus the parameter $b_\nu$. 
The solid and broken lines correspond to the cases $\beta_\nu=0$ and 
$\beta_\nu=20^\circ$, respectively.

\vglue.3in

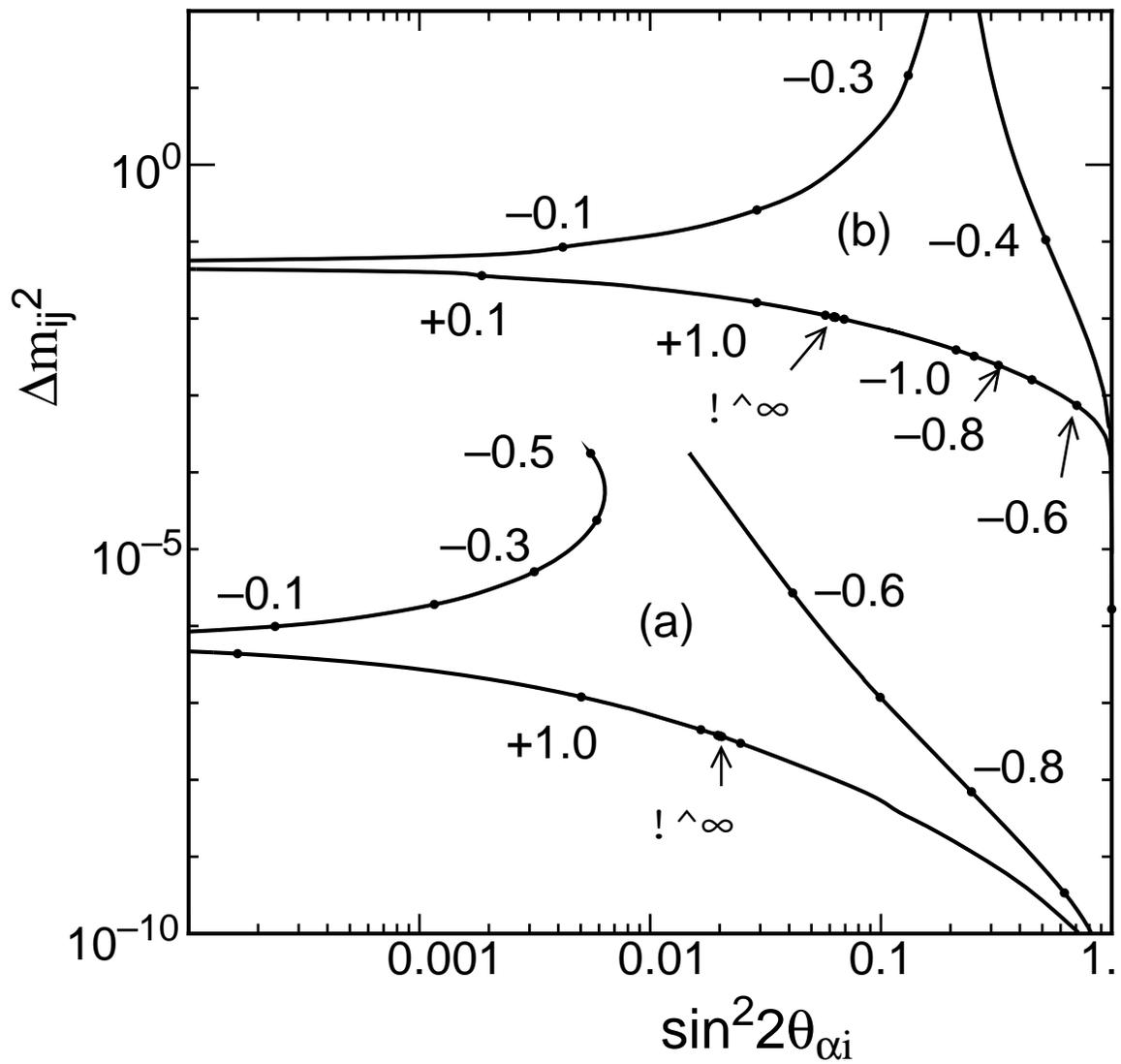
Fig.~2.  $\Delta m_{ij}^2$ [in unit of $(m_0^{\nu})^2$] 
versus $\sin^2 2\theta_{\alpha i}$: 
(a) $\Delta m_{21}^2$ versus $\sin^2 2\theta_{e 2}$ and 
(b) $\Delta m_{32}^2$ versus $\sin^2 2\theta_{\mu 3}$. 
The dots denote points $b_\nu=+10$, $+1$, $+0.1$, $-0.1$, $-0.2$, 
$-0.3$, $-0.4$, $-0.5$, $-0.6$, $-0.7$, $-0.8$, $-0.9$, $-1.0$ and $-10$. 

\vglue.3in

Fig.~3. $\sin^2 2\theta_{e2}$,  $\sin^2 2\theta_{\mu 3}$ and 
$\Delta m_{32}^2/\Delta m_{21}^2$ versus the parameter $b_\nu$. 
The solid and broken lines correspond to the cases $\beta_\nu=0$ and 
$\beta_\nu=20^\circ$, respectively.

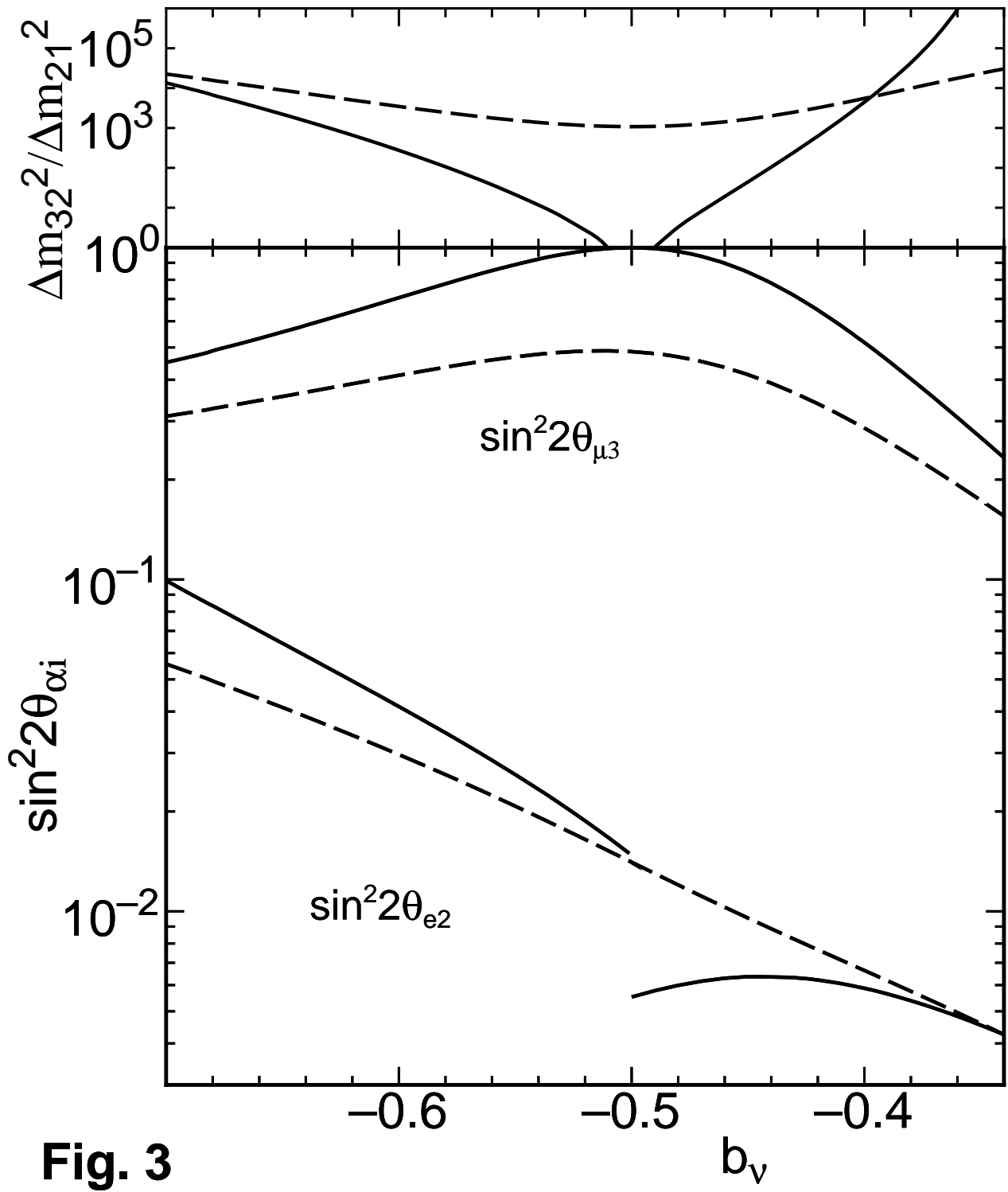
\end{document}